# Template-Assisted Direct Growth of 1Td/in$^2$ Bit Patterned Media

En Yang, Zuwei Liu, Hitesh Arora, Tsai-wei Wu, Vipin Ayanoor-Vitikkate, Detlef Spoddig, Daniel Bedau, Michael Grobis, Bruce A. Gurney, Thomas R. Albrecht, Bruce Terris

HGST, a Western Digital company, San Jose Research Center, San Jose, CA

**Abstract**

We present a method for growing bit patterned magnetic recording media using directed growth of sputtered granular perpendicular magnetic recording media. The grain nucleation is templated using an epitaxial seed layer which contains Pt pillars separated by amorphous metal oxide. The scheme enables the creation of both templated data and servo regions suitable for high density hard disk drive operation. We illustrate the importance of using a process that is both topographically and chemically driven to achieve high quality media.

## Introduction

The difficulties in extending areal densities in magnetic recording media are often depicted as the trilemma of simultaneously achieving high grain density, thermal stability, and write-ability in the media design. Bit Patterned Media (BPM) provides an optimal solution for the grain density problem in the trilemma and is a key enabler for achieving the ultimate theoretically possible magnetic storage areal densities[1]. In BPM, each recorded bit is stored in a single domain magnetic island, rather than a collection of smaller magnetic grains[2]. However, there are numerous media fabrication challenges[3-5] that limit BPM media quality, cost effectiveness, and ease of recording system integration.

The traditional route of creating BPM involves etching continuous magnetic recording media to form the magnetic islands. The etch mask that defines the magnetic islands is created via nanoimprinting using a master template that is reused to make many subsequent disks. The difficulties of etching high aspect ratio grooves, as well as etch mask deterioration and etch damage to the island edges, limit the magnetic fill factors and island heights in etched BPM and reduce achievable performance. In addition, the recording system requires servo patterns for maintaining recording head positioning and etched servo patterns can perturb flying behavior of the recording head. Templated growth BPM media has been recently proposed as a method to create high fill factor magnetic islands, with many other advantages[6]. However, creating high quality single domain BPM islands and servo patterns using templated growth is difficult, in part due to the mismatch in length scales between the as-sputtered magnetic grains and the BPM features.

In this paper, we introduce an epitaxial growth method for fabricating 1.0 Teradot/in$^2$ templated BPM that simultaneously creates both high quality magnetic islands and servo patterns on the same disk. We use nanoimprint lithography to create the templating pattern. The growth template consists of epitaxial magnetic seed layer dots separated by oxide, which facilitates good BPM magnetic island growth and segregation during sputtering of magnetic material. We illustrate the importance of epitaxial growth by comparing magnetic islands grown on epitaxial and non-epitaxial seed features. The magnetic film growth contrast between epitaxial seed and the oxide provides a route to create coexisting servo patterns with the data islands. We demonstrate the quality of the media by recording at better than 1e-2 bit error rate.

## Results and Discussion

The creation of templated growth BPM (TG-BPM) involves directing the grain nucleation processes in conventional perpendicular recording media (PMR) to create segregated single-domain magnetic grains only at desired locations. The templating pattern on the media, which will establish the locations of the grain nucleation sites, can be generated by using "top down"[4] nanolithography or by "bottom up" self-assembled nanostructures[7-12]. In our work the templating pattern consists of carbon or Pt pillars generated using a nanoimprint lithography with a 1.0 Td/in$^2$ template. The template contains features to pattern both the 27 nm pitch hexagonal array of data islands, as well as servo and auxiliary features required by the recording system. NiTa, NiW, and Ru underlayers are deposited on top of the templating layer, prior to depositing the magnetic recording layer. The underlayers and growth parameters are picked to ensure that the top Ru layer is granular with grain locations defined by the template. The magnetic layer is then deposited by co-sputtering of CoCrPt magnetic alloy and oxide segregate materials, similar to those used in conventional perpendicular recording media. The process flow for TG BPM is shown in Fig.1(a).

This novel scheme for creating BPM media has several advantages over etching continuous films of CoCrPt. TG BPM has a much higher fill factor than etched media, as the segregant fraction can be much lower than the magnetic volume fraction removed or damaged by etching. The magnetic layer thickness in TG BPM films can be much larger, as it is not constrained by etching dynamics of high aspect ratio features. Moreover, TG BPM has a continuous top surface, which provides a more stable head-media interface, offers better corrosion resistance, and does not require planarization. The lack of planarization, as well as potentially less stringent patterning requirements, can make TG BPM more cost effective than etched BPM.

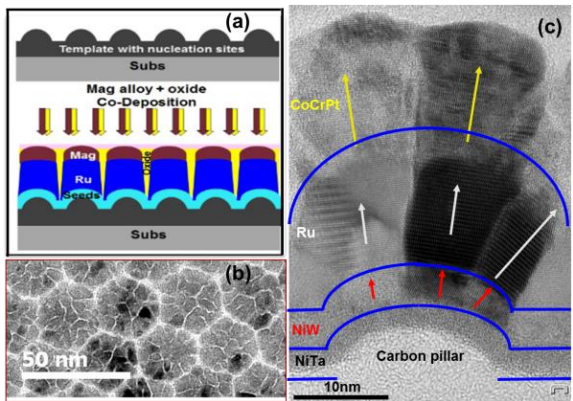

Fig.1 (a) Process flow for templated growth Bit Patterned Media. High resolution Plan view (b) and cross-section(c) TEM images of NiTa/NiW/Ru/Mag+oxide layers deposited on amorphous carbon pillars.

However, there are many challenges associated with fabricating TG BPM with good magnetic properties and recording performance. The templated nucleation sites compete with other grain nucleation and segregation processes, which can cause sub-grain growth, oxide boundaries within a grain, or inconsistent segregation. Insufficiently thick intra-granular grain boundaries can produce undesired and inconsistent exchange coupling between magnetic dots. TG BPM dots need to have high magnetic and structural uniformity in order to have the tight switching field distribution required for recording at a low bit error rate (BER).

The challenge of making high quality TG BPM can be illustrated by using carbon pillars as the nucleation template. Fig.1(b) shows plane-view high resolution transmission electron microscopy (TEM) of the media stack, with substantial sub-grain growth having oxide boundaries within each magnetic dot. The x-ray diffraction rocking curves of the CoCrPt[002] peak show a large 7 degrees full width half maximum, indicating a wide c-axis distribution of the magnetic storage layer that would lead to a wide magnetic switching field distribution. From high resolution TEM cross-section (Fig.2(c)), the cause is clear: the dome shape of the nucleation site (carbon pillar in this case) results in a fan shaped distribution of the c axis of the NiW seed layer, and the fan shaped c axis distribution on each pillar is then epitaxially followed by the sequentially deposited Ru underlayer and the CoCrPt magnetic layer, hence the c-axis rocking angle of the magnetic layer is large, as is its switching field distribution.

While the 1 Td/in$^2$ templating pattern does transfer a pattern to the media, the emergence of widely oriented and isolated subgrains must be eliminated to obtain media with high signal to noise ratio. We have made two key changes that address the aforementioned growth issues. First, the nucleation template pillar material is chosen from among oxidation resistant materials that can be grown with a hexagonal basal plane. Unlike the amorphous carbon pillar, a textured pillar can enable epitaxial growth of fiber textured underlayers with perpendicular c axis orientation. We have successfully used Pt and Pd for the epitaxial nucleation layer material, but other materials with (111) texture, such as fcc Rh, Ir, or Au could also work. Second, we suppress growth of grains outside of the Pt or Pd pillars by introducing chemical contrast via the choice of material in the trench region between the pillars. This is achieved by choosing an easily oxidized seed layer material under the nucleation template layer and etching the Pt pillars all the way to the oxide seed layer. In our case the oxidizing seed layer is NiTa/NiW, which converts to TaOx and WOx during processing.

Fig.2(a) shows a schematic diagram of the process flow for our proof-of-concept TG BPM sample. First, 50nm NiTa, 5nm NiW, and 5nm Pt are deposited on regular glass in a Circulus M12T sputtering system at elevated temperature. Both NiW and Pt are used here to encourage fcc (111) texture with a tight [111] orientation distribution. The fcc (111) close packed basal plane provides a good lattice match with the hcp (0002) plane of Ru and Co magnetic alloys. Pt is also an anti-corrosion material, chosen to survive oxidation from environment as well as subsequent etching processes[13]. NiTa as an amorphous seed layer, which not only provides a smooth surface for better thin film growth quality, but is also easily oxidized. It is designed so that the NiTa will be oxidized during etching/hard mask removal process, or simply via exposure to air. Elevated temperature is used here to promote surface diffusion in order to have a tight [111] orientation distribution, as well as a large grain size for the Pt material.

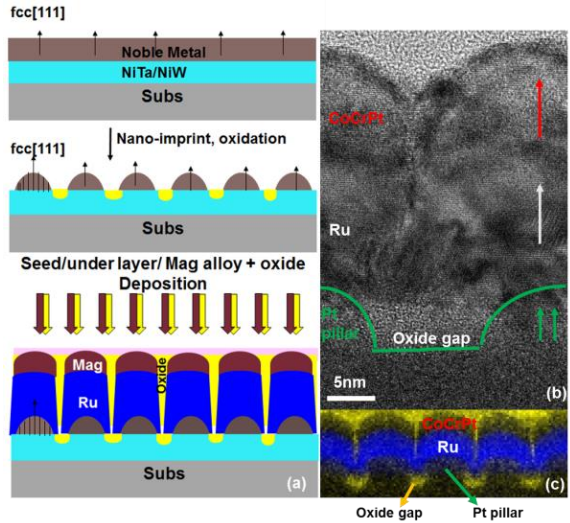

Fig.2 (a) Process flow for epitaxial TG BPM sample, and high resolution cross-section TEM image (b) of a TG-BPM sample and cross-section EELS element mapping (c) for Ru (blue), and oxide(yellow) for the same sample.

The substrates with multi-seed-layers are then patterned by nanoimprinting and IBE/ICP etching processes[9]. The final template contains ~5nm tall Pt pillars, surrounded by oxidized NiTa/NiW gaps (trenches). A 9nm Ru layer, followed by cosputtering of magnetic CoCrPt alloy and oxide, and a 3nm carbon over coat are then deposited on the growth template. In between Pt pillars, the intentionally oxidized NiW/NiTa gaps can be clearly identified in the electron energy loss spectroscopy (EELS) element mapping image in Fig. 2(c). It was found that columnar growth of Ru islands with obvious lateral separation is critical for subsequent phase separation of the magnetic alloy and oxide; these oxide gaps in between Pt pillars can effectively prevent the lateral merging of Ru columns during deposition. Moreover, it is crucial for BPM servo patterns, which will be explained later in this paper. The columnar growth of the Ru layer is clearly revealed in EELS element mapping. On top of Ru layer, the co-sputtered magnetic alloy and oxide are well separated following the guidance of the underlying Ru columns; magnetic alloy grows on the Ru islands, while oxide fills the gaps between Ru islands in order to minimize the total interfacial surface energy of the media.

The cross section high resolution image of TG BPM sample in Fig. 2(b) indicates excellent lattice matched epitaxial growth of close packed basal planes from Pt to the Ru and CoCrPt magnetic alloy. The high quality epitaxial growth was confirmed by x-ray diffraction and magnetic alloy [002] rocking angle measurements. The FWHM of CoCrPt [002] is improved from 7 degrees to ~2.5 degrees, indicating a much tighter c axis orientation distribution of magnetic recording layer from the new process.

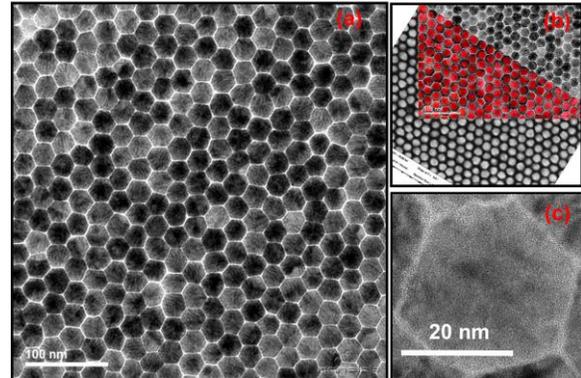

Fig.3 (a) High resolution plan-view TEM image of the TG BPM sample (b) Overlay (center) of a SEM image of the growth template (left down) with Pt pillars and the TEM image (up right) of the same sample after recording media deposition. (c) TEM detail of one hexagonal magnetic island.

Fig.3 (a) shows a TEM plan-view image of the improved TG BPM. Hexagon shaped magnetic islands are uniform both in size and shape. From the superposition of the TG BPM TEM image and the growth template SEM image before deposition shown in Fig.3 (b), it can be clearly seen that the magnetic islands are well registered with the underlying Pt nucleation pillars. Further image analysis confirms that these two images have the same, tight pitch distribution. Moreover, the irregular shape and non-uniform size of the Pt pillars in the template has grown into a uniform hexagon shape after film deposition. A high resolution TEM image of one island shown in Fig.3(c) further reveals a CoCrPt (0001) plane top view lattice structure; each magnetic island is well defined and isolated by oxide, with no sub-grains or oxide grain boundaries are observed within one island, and no obvious bridges to the neighbor island as well.

Fig.4 shows a direct comparison of the out-of plane polar Kerr Loops for TG BPM samples from the improved new method (on Pt pillars with oxide gap) and previous non-epitaxial growth (on carbon pillars) The epitaxial sample loop has excellent perpendicular orientation, high coercivity (~8 kOe), high negative nucleation field (-5.9kOe), and high thermal stability, with KuV/KT>270. Those values are similar to theoretically calculated values for ideal 25nm diameter 7.8nm tall hexagon shaped CoCrPt islands with the given composition. Furthermore, the intrinsic switching field distribution (iSFD/Hc) is improved from 50% to 11% by using new epitaxial method. The iSFD can be further improved by using an exchange coupled composite structure in the magnetic recording layer[14].

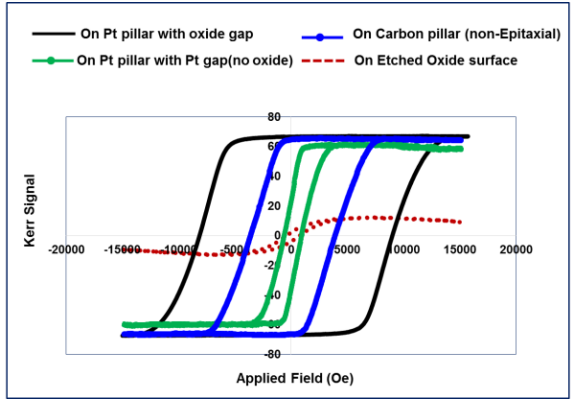

Fig.4 Magnetic hysteresis loops of 9nm Ru\7.8nm CoCrPt+oxide deposited on four different surfaces.

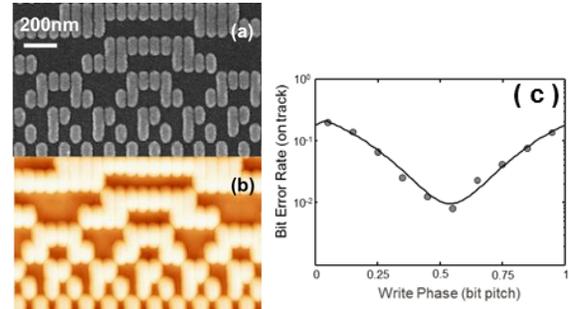

Fig.5 (a) SEM image of a BPM servo sector pattern on a sample produced by templated growth; (b) MFM image showing magnetic readback signal from a region with the same pattern (c) Graph of bit error rate versus write phase.

We confirmed the high quality of the TG BPM media by performing bit error rate measurements using a drag tester with commercially available HDD heads. The details of our experimental setup can be found in reference.[15] On-track recording using pseudo-random bit sequences shows a bit error rate of 1e-2 when the switching of the write current polarity relative to the island locations is optimized (Fig. 5c). The bit error rate increases when the write phase deviates from optimal, indicating that the BPM islands are magnetically indivisible. While lower bit error rates have been achieved with etched BPM[6], the generally good recording performance of TG BPM shows significant promise.

One proposed mechanism for the templating effect of the nucleation seed is shadow growth.[16] In this model the topographic bumps in the nucleation seed layer are amplified during subsequent deposition steps to create a granular structure in the underlayer. The ability of amorphous carbon bumps to template magnetic grains provides some support for this mechanism. The presence of additional nucleation sites on the carbon pillar templated films indicates that other nucleation mechanisms may be active which are reduced when the carbon pillars are replaced with epitaxial Pt separated by oxide. To test whether the additional chemical contrast plays an import role in the growth dynamics, we grew the same film stack on dense Pt pillars without oxide gaps. As shown in Fig. 4, the Kerr loop shows poor magnetic properties with low coercivity, indicating large exchange coupling between magnetic islands. The poorer magnetic quality indicates that shadow growth alone cannot account for the templating effect and segregation guidance through chemical contrast is an important element of good templated grain growth. It is worth noting that the TG BPM growth process can also be used to grow non-hexagonal arrays of grains, enabling high bit aspect ratio rectangular cells to be grown in this manner.

In addition to highly uniform data regions, the hard disk drive (HDD) storage system requires complex sector header and servo positioning patterns on the disk to enable normal drive operation. The readback information from these patterns provides position and timing information, without which the recording system cannot locate, read, and write user data correctly. Typically, there are several hundred servo regions located at fixed intervals on every data track. Etched BPM offers the possibility of creating servo regions through patterning, rather than servo writing, which saves significant time and cost during drive manufacturing. Fig. 5(a) shows an SEM image of a section of the servo region in a TG BPM disk. If this were an etched BPM disk, the light grey regions would contain magnetic material, while the dark regions would be areas from which the magnetic material has been etched away. Following DC magnetization, the readback signal is generated by the presence or absence of magnetic material. A noticeable feature in the servo region is the series of progressively smaller "falcon wing" patterns which are part of a Gray code encoding scheme for denoting track numbers. The patterns are further subdivided into smaller islands to limit domain wall formation within the extended features. However, even these smaller islands can be two to four times larger than the data islands. Creating pre-patterned servo patterns in TG BPM has several challenges. First the servo pattern feature sizes are significantly larger than the data islands and can conflict with the natural segregation and nucleation lengths of the magnetic stack. As a result the magnetic quality of the servo features can be significantly worse than the data islands. Second, magnetic material is still deposited in the large regions between the templated nucleation regions. The non-templated regions can thus contain magnetic grains that are similar to the templated grains[17]. As a result, DC magnetization might not produce adequate magnetic contrast for servo readback, defeating the purpose of pre-patterning the servo.

The key to creating useable pre-patterned servo in TG BPM is to make the non-templated regions incapable of supporting growth of perpendicularly oriented magnetic media. If in addition suitable magnetic grains can be grown on the larger patterned Pt(111) servo features, then a patterned Pt template could be used as a growth seed for the servo area. Similar to etched BPM servo, the TG BPM servo would be DC magnetized and the readback signal arising from magnetic contrast between the well-ordered and poorly ordered regions.

In Fig. 4, the red (dot) curve shows that etched oxide surface (oxidized NiTa/NiW) is incapable of supporting growth of high quality perpendicularly oriented media, which can be used as an excellent growth seed in the servo non-templated regions. The oxidized NiTa/NiW underlayer, which enhances segregation in the data regions, provides an amorphous non-epitaxial surface for growing non-textured Ru and hence randomly oriented CoCrPt in the servo trench regions. Moreover, the additional roughness resulted from etching process leads to smaller grain growth, shorter diffusion length and higher Cr contents in the magnetic film, which further reduces magnetic signal and magnetic anisotropy, ensures that oxidized NiTa/NiW surface is suitable for creating low magnetic signal Servo trenches.

Fig. 5(a) shows an SEM image of a TG BPM sample after 9nm Ru and 8nm magnetic alloy+oxide multilayer deposition. Fig.5 (b) shows an MFM image of the same region, providing magnetic servo readback signal of the same region. It can be seen that with strong magnetic contrast between templated and non-templated regions (features and trenches), the MFM image shows the same pattern with magnetic signal as seen in the SEM image.

**Conclusion**

We have demonstrated the ability to make high quality 1Td/in$^2$ BPM using templated growth of sputtered films on a surface of Pt pillars isolated by oxide trenches. The process is capable of generating servo and data regions on the same disks. We have illustrated the importance of chemical and topographic contrast in the template for creating high quality magnetic islands. Recording measurements on the TG BPM data islands show a 1e-2 bit error rate, indicative of high quality magnetic islands with excellent indivisibility.

*Conflict of interest:* The authors declare no competing financial interest.